\documentclass[nohyper,12pt,letterpaper]{JHEP}  
\usepackage{epsfig}

\newfont{\frak}{eufm10 scaled 1200}

\newfont{\Bbb}{msbm10 scaled 1200}     
\newcommand{\mathbb}[1]{\mbox{\Bbb #1}}
\DeclareSymbolFont{AMSa}{U}{msa}{m}{n}
\DeclareSymbolFont{AMSb}{U}{msb}{m}{n}
\let\Box\relax
\DeclareMathSymbol{\Box}{\mathord}{AMSa}{"03}

\def \eqn#1#2{\begin{equation}#2\label{#1}\end{equation}} 

\def\hacek{\accent20}                           

\title{Remarks on M Theoretic Cosmology}

\author{Tom Banks\\
  Department of Physics and Astronomy\\
  Rutgers University, Piscataway, NJ 08855-0849\\
E-mail: \email{banks@physics.rutgers.edu}}

\abstract{I present cosmological arguments which point towards a
Horava-Witten like picture of the universe, with the unification scale of
order the fundamental gravitational scale.  The SUSY breaking scale is
determined by the dynamics of gauge fields which are weakly coupled at
the fundamental scale.  Bulk moduli whose potential originates at short
distances are the inflatons, while bulk moduli whose potential
originates from SUSY breaking are the origin of the energy density in
the present era.  The latter decay just before nucleosynthesis, and a
consistent theory of baryogenesis requires that there be renormalizable
baryon number violating interactions at the TeV scale.  The dark matter
is a boundary modulus, perhaps the QCD axion, and the temperature of
matter radiation equality is related to the ratio between the
fundamental and effective four dimensional Planck scales.  The same
ratio determines the amplitude of fluctuations in the microwave background.

}


\received{???????? ?st, 1998}
\accepted{???????? ?th, 1998}

\preprint{\hepth{9906126}\\RU-99-22}
\begin{document}

The purpose of this note is to revisit certain issues in string 
(now M theory) cosmology, in
light of the string duality revolution.  Our previous investigations of 
these issues are 
contained in \cite{preva}\cite{bdaxion}.  Our basic claim is that
certain observations in cosmology
suggest (though we will certainly not claim that they prove) that the
world is close to
a vacuum state of M-theory similar to that which arises from Calabi Yau
compactification of
the strongly coupled heterotic string of Horava and Witten \cite{horwita}.  
We will describe what may be a large class of generalizations of the
Horava Witten scenario.

The major issues we discuss may be itemized as follows:
\begin{itemize}

\item One of the central claims of \cite{preva} , following the seminal
work of \cite{binetry}
, was that moduli of string theory are natural inflaton candidates.
However, there is
a sort of contradiction in this statement.  Moduli are cleanly defined
only when they are
true zero modes.  Inflatons must have a potential.  In \cite{preva} this 
contradiction was
hidden beneath the assumption that the natural scale of the potential
during inflation is
much smaller than the fundamental scale of string theory, which was 
identified with the
four dimensional Planck mass.  However these papers also made much of 
the coincidence
between the scale of inflationary vacuum energy and the Unification
scale.  Witten's
explanation of the ratio between the Unification and Planck scales in 
the context of
the Horava-Witten scenario\cite{horwitb}, identifies the former with the
 fundamental scale of the
theory.  In this context, the separation of inflaton fields with a
potential of order
the fundamental scale of M theory from the rest of the high energy
degrees of freedom
of the theory seems highly suspicious.  
We will see that these conundra are simultaneously resolved in any
scenario with
8 supercharges (SUSYs) preserved in the bulk and only 4 preserved on 
certain lower dimensional
submanifolds or branes.  We will argue that there may be many such
compactifications of
M theory, generalizing that of Horava and Witten.  In such 
compactifications, $N_e$ e-foldings
of inflation require only a small coefficient of order $1/N_e$ in 
the superpotential of
bulk moduli.  The size of primordial energy density fluctuations is naturally 
explained in terms of the unification scale and the four dimensional 
Planck scale.

\item In \cite{preva} the existence of an independent, lower, scale for 
SUSY breaking was
considered an embarrassment.  Here we ascribe it to the existence of a
true moduli space of
the low energy effective theory with 4 SUSYs, which is identified as 
the locus of an enhanced
discrete R symmetry with certain properties.  Nonperturbative low energy gauge 
interactions spontaneously break both this R symmetry and the remaining 
four supercharges.
The details of the resulting physics depend somewhat on whether we 
assume the true moduli
space contains bulk moduli, or only fields whose origin is on one 
of the branes.
We somewhat prefer the case with bulk moduli because it leads to a 
natural explanation of
the relation between the scales of R symmetry and SUSY breaking, 
that is necessary in any
supergravity theory in which the cosmological constant vanishes.  
In passing we note that
this scenario for SUSY breaking resolves the overshoot problem 
of \cite{brustein}.

\item Both bulk and brane modes of SUSY breaking lead to a version of 
the cosmological
moduli problem \cite{othersandbks}.  We review a resolution of this 
problem \cite{bdaxion}
in which the moduli decay and reheat the universe to a temperature just 
above that needed
for nucleosynthesis.  The baryon asymmetry is produced in the decay of 
the moduli.
This implies that the low energy theory at energies of order 1 TeV 
contains renormalizable baryon number violating interactions .   As a 
consequence, there
is no natural SUSY candidate for dark matter.

\item We review and generalize the observation of \cite{bdaxion} that 
a brane modulus with
sufficiently small potential is a natural dark matter candidate in the 
above scenario.
The QCD axion (with a decay constant of order the unification scale) is
a possible realization
of this mechanism.  However, there is nothing which requires 
the potential of the 
dark matter candidate to be of the QCD scale.

\end{itemize}

\section{Moduli and Inflation}

Generic compactifications of M theory to four dimensions with only 
four SUSYs have no moduli.
That is, there are no theorems which prevent the occurence of 
a superpotential on the would be 
moduli space.  Furthermore, all symmetries of M theory are gauge 
symmetries.  As a 
consequence
of D-terms, the true moduli space is made up of fields invariant 
under all continuous 
gauge symmetries.  Thus, the superpotential is restricted only by 
discrete symmetries
and these generically do not require it to vanish.  
The exception is a discrete complex R symmetry, \cite{bdmod}.  The 
submanifold of field space
invariant under such a symmetry, and containing no directions of R charge 2,
 is a true moduli space\footnote{Another possible region of 
field space where the superpotential vanishes has been explored by 
Witten \cite{witsup}.  
Witten's argument for vanishing superpotential uses a $U(1)$ symmetry 
valid only in 
a certain large volume limit to draw exact conclusions about 
the superpotential.  I find it
mildly suspicious and will not include Witten's region in the 
discussion here, but it may
have an important role to play.}.  

Most discussions of the phenomenology of string/M theory have 
been based either on 
low energy 
SUGRA, or weakly coupled string expansions.  In these discussions 
the apparent moduli 
space of
the theory is much larger than the true moduli space.  There are
theorems which prevent the
occurence of a superpotential to all orders in the perturbative 
expansion.  If one works in
the regime where nonperturbative corrections to the superpotential 
are small then the phrase
\lq\lq superpotential for moduli \rq\rq is not an oxymoron.  It was in 
this context that
the idea of moduli as inflatons was 
proposed \cite{binetry}\cite{preva} .  A serious
problem with this regime was pointed out long ago by Dine and 
Seiberg \cite{dineseib}.
Within the context of a systematic perturbative expansion one 
cannot stabilize the moduli
(small couplings or large radii) whose large values justify the 
expansion.  Racetrack models
\cite{racetrack} and Kahler stabilization \cite{bdcoping} are 
two attempts to get around this
problem.  Neither leads to a reliable calculational framework, and 
their fundamental 
postulates have not been verified.  

Another reason for avoiding extreme regions of moduli space was 
pointed out by Moore and Horne
\cite{moore}.  In extreme regions of moduli space, the metric on the 
space can be reliably
calculated and the infinite regions have finite volume.  This means that 
the system 
dynamically
avoids the extreme regions.  For example, if the potential has two
minima with vanishing
cosmological constant, one in the interior of moduli space and the other 
in one of the
extreme regions, then a generic motion of the system will end up at the 
minimum in the 
interior.

In a vacuum state with no large moduli on the other hand, it 
is not clear what the term
modular inflation could mean.  This is particularly true if one 
adopts Witten's explanation of
the ratio between the Planck and unification scales, with the 
concomitant conclusion that
the fundamental scale of M theory is on the order of $10^{16}$ GeV.  
The simplest interpretation
of the magnitude of primordial energy density fluctuations in 
inflationary cosmology
invokes a vacuum energy during inflation of approximately this 
order of magnitude.  In what sense
can fields with a potential energy of this order of magnitude 
be considered moduli?  
Recall that the motivation for separating moduli out from 
the other degrees of freedom of M theory
is that they are supposed to parametrize low energy motions of 
the system among would be
ground states.

In fact, scenarios like that of Horava and Witten contain the 
clue to an answer to this
question.  The universe is separated into \lq\lq branes \rq\rq and 
bulk, and the latter has
more SUSY than the former.  Thus, the bulk universe has 8 
approximately conserved
supercharges and thus contains fields which would act as true 
moduli if it were not for the 
presence of the branes.  The superpotential for the moduli 
is generated on the branes.

Let us examine the consequences of this fact.  At an energy scale 
small compared to the 
mass of the bulk
Kaluza-Klein modes on the compact manifold, the world is effectively 
four dimensional.
The moduli become fields in this four dimensional effective 
field theory.  Since the 
effective
theory has only four SUSYs, these fields have a superpotential.  
Since it comes only from the
vicinity of the branes on which the larger SUSY algebra is broken, 
it is independent of the
volume of the internal space, and has, by dimensional
analysis, the form
\eqn{supot}{W = M^3 w(\theta_a )} 
where $\theta_a$ are dimensionless parameters characterizing 
the internal geometry.
On the other hand, the kinetic term for these zero modes, just 
like the Einstein term
for the zero modes of the gravitational field, is proportional to 
the volume $V_7$ of the 
internal manifold, and has the form
\eqn{kin}{M^9 V_7 \sqrt{-g} G_{ab} (\theta ) \nabla \theta_a \nabla \theta_b .}
Note that $M^9 V_7 = m_P^2 = {1\over 8\pi G_N}$ is, 
as the notation indicates, the same
coefficient which multiplies the Einstein action.  Furthermore, although 
the volume
$V_7$ is itself a modulus, when we pass to the Einstein conformal frame 
in which $V_7$ is replaced 
by its vacuum value, the kinetic term of the moduli is rescaled 
in precisely the same manner
as the gravitational action.   It is then natural to define canonical
scalar fields 
by $\phi_a = m_P \theta_a$.  Their action has the form
\eqn{action}{\int \sqrt{-g} G_{ab} (\phi /m_P ) 
\nabla \phi^a \nabla \phi^b - {M^6 \over m_P^2} v(\phi /m_P ).}

The slow roll equations of motion derived from this action are
\eqn{slowroll}{3H d \phi^a /dt = - {M^6\over m_P^2} G^{ab} 
{\partial v \over \partial\phi^b}.}
and lead to the equation
\eqn{vdot}{d v /dt = {M^3 \over 3 m_P^2 \sqrt{v}} 
\partial_a v G^{ab} \partial_b v.}
where $\partial_a$ refers to the derivative with respect to the 
dimensionless variable $\theta^a$.
We have also used the slow roll expression for $H$ in terms of the potential.
{}From \ref{vdot} we immediately derive an expression for the number 
of e-foldings
\eqn{efold}{N_e = 3 \int {v\over \partial_a v 
G^{ab} \partial_b v} \partial_c v d\theta^c    .}
where the integral is over the trajectory in moduli space that 
the system follows
during the time interval when the slow roll approximation is valid.  
We see that in order to
obtain a large number of e-foldings we need a potential which is 
flat in the sense that
$|\partial v|/v \sim 1/N_e$.  The phenomenologically 
necessary $N_e \sim 60$ can be achieved with
only a mild fine tuning of dimensionless coefficients.  
Correspondingly, the conditions on the
potential which ensure the validity of the slow roll approximation 
are order one conditions on
the derivatives of the potential and do not contain any small 
dimensionless numbers.

The fact that actions of the form \ref{action} give rise to 
inflation with minimal fine tuning, 
and that such actions naturally arise for moduli in string theory 
was pointed out in \cite{preva}.
The general point that moduli might provide the flat potentialled, 
weakly coupled fields 
necessary to inflation was first made in \cite{binetry}.  Here we note 
that in brane scenarios,
it is the {\it bulk moduli} which play this role.  By contrast, moduli 
associated with a single
brane will have a natural scale $M$ and do not play the role of 
inflatons in quite as gracious
a manner.

Another pleasant surprise awaits us when we plug the potential 
from \ref{action} into the
standard formula for the amplitude of the primordial energy density 
fluctuations generated
by inflation.  Up to numbers of order one we find
\eqn{deltarho}{{\delta\rho \over \rho} \sim N_{\lambda} (M/m_P)^3 \sim 10^{-5}}
where the numerical value comes from the measured cosmic microwave
 background fluctuations, and $N_{\lambda} \sim 50$.
This gives $M \sim (2/10)^{1/3} \times 2 \times 10^{16}$ GeV, which, 
given the crudeness of the
calculation, is the unification scale.   To put this in the most 
dramatic manner possible, we
can say that a brane scenario of the Horava-Witten type, given 
the unification scale as
input, {\it predicts the correct amplitude for inflationary 
density fluctuations}.  Furthermore,
the whole scenario only makes sense because of the same large volume 
factor that underlies
Witten's explanation of the ratio between the Planck and unification 
scales.  This is necessary 
at a conceptual level to understand why it is sensible to think 
about a modulus with a super
potential of order the fundamental scale, and at a phenomenological 
level to understand the
magnitude of the density fluctuations.  

Although it has no connection with our discussion here we cannot 
resist pointing out the other
piece of evidence for a scale of the same order as $M$.  Any theory of 
the type we are discussing
would be expected to contain corrections to the standard model Lagrangian 
of the form (in superfield notation) ${1\over M} L L H^2 $, which gives 
rise to neutrino masses.
It is a matter of public record \cite{superK} now that such masses 
exist, with an estimated value
for $M$ between $.6$ and $1.8 \times 10^{15}$ GeV.  Although this 
is an order of magnitude shy
of the unification scale I believe the uncertainties in 
coefficients of order one in dimensional
analysis could easily make up the difference.  If not, we will have 
the interesting problem
of explaining the existence of two close but not identical energy 
scales in fundamental physics.
\cite{wilczeketal}.  

Finally, we want to note that this scenario for inflation does not 
suffer from the runaway problem pointed out by Brustein and Steinhardt 
\cite{brustein}.  These authors noted that the inflationary vacuum energy
is much larger than the SUSY breaking scale.  Furthermore, the minimum of
the effective potential was assumed close to the region of weak string
coupling.  There was then a distinct possibility that the inflaton field would
overshoot the small barrier separating it from the extreme weak coupling
regime where string theory is incompatible with experiment.  In the present
scenario, the coupling is not assumed to be weak (nor the volume extremely
large).  Furthermore the inflationary potential has nothing to do with
SUSY breaking.   There is no runaway problem at all.

\subsection{SUSY breaking}

The authors of the papers in \cite{preva} agonized over 
the discrepancy between the
unification scale and the scale of SUSY breaking.  In fact, they 
discussed and discarded what
I now believe is the obvious solution of this problem, because 
of problems specific to weakly
coupled string theory.  The obvious way to avoid SUSY breaking 
at the scale $M$, is to
insist that the superpotential \ref{supot} has a SUSY minimum.  
In fact, the existence of
such minima is  generic
, requiring only the solution of $n$ complex equations for $n$ unknowns.  
However, in general, the superpotential will not vanish at such 
a minimum but instead 
will give rise to a negative cosmological constant.  We refer the 
reader to \cite{preva} for the
elementary argument that in a postinflationary universe, such a 
SUSY point in moduli space
is not a stable attractor of cosmological solutions.  Instead, 
generic solutions which try to
fall into such a minimum, recollapse on microscopic time scales.  

The stable postinflationary attractors of a supersymmetric cosmology 
are points in moduli space
with vanishing superpotential and SUSY order parameters.  These can 
be characterized in terms
of a symmetry.  Namely, any complex R symmetry forces the 
superpotential to vanish, and if there
are no fields of R charge 2 then the SUSY order parameter 
vanishes as well.  The R symmetry must
of course be discrete, since we are discussing M theory.  If in 
addition, there do exist
fields of R charge 0, then there will be an entire submanifold on 
which the superpotential
vanishes and SUSY is preserved.  Our future considerations will 
concentrate on this submanifold, 
which, following the terminology in the introduction, we call the 
true moduli space.
It is the locus of restoration of a discrete R symmetry with the 
above properties.

Before proceeding to the discussion of SUSY breaking on the true 
moduli space, we should
introduce the final characters in our story, the boundary or 
brane moduli.  We could in fact have
inserted such fields, which arise as excitations localized on one 
of the branes, into
our discussion of inflation.  However, they would have been of 
little use there, as their
natural scale is $M$ rather than $m_P$ and they are rapidly driven to
their instantaneous minima during
the inflationary era.  At lower energies however they will play 
an interesting role.

In addition to these moduli fields, any brane scenario will 
contain a variety of gauge fields and
matter fields in nontrivial representations of the gauge group.  
The moduli will interact with
these fields via the moduli dependence of bare gauge and yukawa 
coupling parameters in the effective theory as well as thru a variety of 
irrelevant operators.  If the gauge couplings are asymptotically 
free and do not run to
infrared fixed points at low energy, this description of the
physics only makes sense if the bare gauge couplings are sufficiently 
small that the scale 
at which the effective 
coupling becomes large is substantially below the scale $M$.  
Otherwise it is not
consistent to include the gauge degrees of freedom in the low energy 
effective theory.
The weakness of bare couplings in these scenarios is not evident a 
priori, as it would be 
in a purely perturbative approach.  The underlying physics is assumed to 
be strongly coupled.
Witten has shown how the small unified coupling of the standard model 
can be explained in terms
of a product of a large number of factors of order one in a geometry 
of large dimensions.
We will assume that similar numerical factors explain the strength of
the gauge interactions that lead to SUSY breaking.  

The main role of the gauge interactions is not to break SUSY, but 
rather the discrete R
symmetry.  If we fix the moduli and treat the gauge theory as a 
flat space quantum field
theory, then SUSY remains unbroken even though a nonperturbative 
superpotential is generated.
The scale of this superpotential is determined via a standard 
renormalization group
analysis in terms of the bare gauge coupling function 
$f(\phi /m_P , \chi / M)$, where we 
have indicated dependence on both bulk and boundary moduli.  
For simplicity we assume that
$f$ is a large constant $f_0$ plus a smaller, moduli dependent, 
term.  The conclusions are 
not affected by this assumption.  The scale $\mu$ of the 
nonperturbative superpotential
is then determined by $f_0$.  It takes the form
\eqn{supottwo}{W_1 = \mu^3 w_1 (\phi /m_P , \chi /M)}
We have eliminated all (composite) superfields related to the 
gauge interactions from this
expression by solving their F and D flatness conditions for fixed 
values of the moduli.
The possibility of doing this is equivalent to the statement that 
the gauge theory does not itself break SUSY.
We assume that $W_1$ does not vanish at any minimum of the effective potential.
This is the statement of spontaneous R symmetry breaking.  
As a consequence, SUSY minima of the potential have negative 
cosmological constant of
order at least $\mu^6 / m_P^2$ and are not attractors of the 
cosmological equations.
Thus, cosmologically, R symmetry breaking forces the moduli to choose 
a minimum with 
spontaneously broken SUSY\footnote{The tunneling amplitudes of such 
nonsupersymmetric
vacua into supersymmetric AdS vacua are incredibly tiny and might be 
identically zero, 
as discussed in \cite{preva}.}.  

Phenomenology requires a value of $\mu$ which gives acceptable squark 
masses.  The details depend
on whether or not we can set the F terms of the boundary moduli equal to zero 
(if there are no bulk
moduli this is not consistent with our other assumptions).  If we can, 
then the nonvanishing
F terms are of order ${\mu^3 \over m_P}$.  A standard argument shows 
that squark masses
will be of order $ {\mu^3 \over m_P^2}$, about the same as the gravitino.  
Assuming this is about a TeV we find $\mu \sim 10^{13}$
GeV .   An attractive feature of this scenario is that the positive and 
negative terms
in the SUGRA potential are naturally of the same order of magnitude.  
Although we have no
real understanding of why the cosmological constant is so small, this 
fact of nature is an
indication of a relation between the scales of R symmetry breaking 
and of SUSY breaking.
In models in which the SUSY breaking F term originates as a bulk modulus the
correct order of magnitude relation between these scales arises 
automatically.  

By contrast, if we assume that the SUSY breaking F term is that of a 
boundary modulus,
the negative term in the potential is of order ${M^2 \over m_P^2} 
\sim 10^{-4}$ smaller than the
positive term.  To understand the cancellation of the 
cosmological constant, one can
, following \cite{preva} introduce two gauge groups.  The first leads 
to spontaneous R symmetry
breaking with unbroken SUSY at a scale $\mu_1$ while the second breaks SUSY at
$\mu_2$.  If $(\mu_1 /\mu_2)^6 \sim m_P^2 / M^2$ one can again 
obtain "order of magnitude
cancellation" of the cosmological constant, but the scenario 
clearly lacks simplicity.  
In this scenario squark masses are of order $\mu_2^3 / M^2$, and 
the gravitino is lighter than 
this by a factor $\sim 10^{-4}$ and weighs about $100 MeV$.  
$\mu_2$ has to be about $5 \times 10^{11}$ GeV.  

The first of these scenarios is clearly simpler, but as we now recall, 
it leads to the
cosmological moduli problem.  The scalar fields in the bulk moduli 
multiplets acquire masses from 
the SUSY violating potential of order $m_M \sim \mu^3 / m_P^2$ which is 
the same order of magnitude
as the gravitino and squark masses, {\it i.e.} a TeV.  They have only 
nonrenormalizable
couplings to ordinary matter, scaled by $m_P$.  Thus, their nominal 
reheat temperature , 
$\sqrt{m_M^3 /m_P}$ is of order $\sim 3 \times 10^{-2}$ MeV, 
and the universe is matter
dominated at the time that nucleosynthesis is supposed to be taking
place.  The thermal 
inflation scenario \cite{therminf} can solve this problem, 
and we will review another solution
\cite{bdaxion} below, but it might tempt us into adopting the 
scenario with boundary moduli
as the instigators of SUSY breaking.  

In this case, one would assume that all bulk moduli
are frozen by the initial superpotential of order $M^3$.  
Dine \cite{mike} has advocated that
the proper vacuum should be an enhanced symmetry point of moduli 
space at which all moduli
(he does not make a distinction between bulk and boundary fields) 
are nonsinglets.  We temporarily
adopt this point of view, but only for the bulk moduli.  
Then the boundary moduli masses are of order $1$ TeV, but their 
couplings to ordinary matter
are scaled by $M$ rather than $m_P$.  The reheat temperature is 
rescaled by a factor of $10^2$ 
and is (just) above the temperature for nucleosynthesis.  
The Hot Big Bang occurs just in time
to light the furnace in which the primordial elements were formed.  

One still has to account for baryogenesis.  Adopting a mechanism 
suggested long ago by Holman, Ramond and Ross
\cite{hrr} we aver that 
this can come from the decay of the moduli themselves.  
All of their interactions
are of order the fundamental scale of M theory, so there is no 
reason for them to preserve
accidental symmetries like baryon and lepton number.  It is quite 
reasonable that they also
violate CP, though the status of CP in M theory is somewhat more 
obscure.  The decay itself
is an out of equilibrium process, so all of the Sakharov criteria 
for baryogenesis are
fulfilled.  However, we must also take note of the theorem of 
Weinberg \cite{steve}, according
to which baryon number violating terms in the Hamiltonian must act 
twice in order to generate
an asymmetry.  In the decay of moduli, the first action of the 
Hamiltonian comes at no cost 
in amplitude, because the modulus must decay somehow and there is no 
reason for its baryon 
number
violating decays to be significantly smaller than those which conserve 
baryon number.
However the second baryon number violating interaction should not be 
highly suppressed if
we want to generate a reasonable baryon asymmetry.  Indeed, a one TeV,
gravitationally coupled,  particle which produces 
a baryon asymmetry of order one in its decay, also produces of 
order $(1 {\rm TeV}/ 3 {\rm MeV})$
or $\sim 3 \times 10^5$ photons.  Thus a large suppression of the 
average baryon number per
decay would give too small a baryon asymmetry.  A way out of this 
difficulty is to admit
renormalizable baryon number violating operators in the supersymmetric 
standard model.
Discrete symmetries such as a $Z_2$ lepton parity \cite{irnir} can 
adequately suppress
all unobserved baryon and lepton number violating processes in the 
laboratory, while allowing
such operators with quite large coefficients.  
An unfortunate casualty of this mechanism is the lightest SUSY particle.  
The LSP is no longer
stable in the scenario described above and we have to look 
elsewhere for a dark matter candidate.

With this scenario in mind, let us return to the situation with 
bulk moduli.  Suppose that
the coefficient in the order of magnitude relation between the moduli 
mass and the fundamental
parameters is $m_M = 5 \times \mu^3 / m_P^2$, while the squark mass is 
actually $m_{\tilde{q}}
= \mu^3 /4m_P^2 = 1$ TeV.  Then the reheat temperature for the bulk
moduli is multiplied by
a factor of $20^{3/2} \sim 10^2$ and is again just above $1$ MeV.  
Nucleosynthesis is again
saved and baryogenesis can take place in the process of reheating.  
Again we must invoke 
renormalizable baryon number violation.  Now however, there are 
natural candidates for
dark matter.  Imagine a boundary modulus whose potential energy is 
substantially smaller
than the the estimate $\mu^3 / M^2$ coming from \ref{supottwo}.  We 
will call this the dark
modulus, because it will be our dark matter candidate. It has a 
potential of the form
$U = \Lambda^4 u(D/M)$. (In \cite{bdaxion}, where
this scenario was first proposed, the candidate was a QCD axion field 
(which arises under
certain natural conditions in Horava-Witten scenarios\cite{strongcoup}).  
This model works, but the mechanism
is much more general and does not require energy densities as small as 
those of the axion.

Now, briefly review cosmic history.  First we have inflation 
generated by bulk moduli fields
which are not on the true moduli space\footnote{Perhaps we 
should call these {\it inflamoduli}.}
.  This period ends after of order $100$ e-foldings,
and the universe is heated by inflamoduli decay to a temperature of 
order $10^9$ GeV .  
The primordial plasma quickly redshifts away.  Furthermore, as soon as 
the inflamoduli
potential energy density falls to $\mu^6 / m_P^2$, the universe 
becomes dominated by the
coherent oscillations of the true bulk moduli.  
The dark modulus remains frozen at some generic point on its potential 
until the Hubble
parameter falls to the mass scale of this field.  At this point 
the energy density of the
universe is of order $\rho \sim m_P^2 \Lambda^4 / M^2$ which is of 
order $(m_P/M)^2 \sim 
10^{4}$ times larger than the energy density of the dark modulus.  
The important point now
is that this ratio is preserved by further cosmic evolution until the 
true bulk moduli decay.
After that time, the dark energy density grows linearly with 
the inverse temperature relative
to radiation, and
matter radiation equality occurs at $10^{-4} MeV$.  This is close 
enough to the true value
for the observable universe that the factors of order one which 
we have neglected throughout 
might account for the difference.   $\Lambda$ must satisfy two 
constraints in order
for this scenario
to work:  the dark moduli must remain frozen until the true bulk 
moduli begin to oscillate, 
and 
the dark modulus must have a lifetime at least as long as the age of 
the universe.
The second constraint is by far the stronger, and leads to 
$\Lambda < 3 \times 10^6$ GeV.
Axions satisfy this constraint by a large margin.  Note that this 
scenario completely
removes the conventional cosmological constraint on the axion 
decay constant.  Axions will
be very weakly coupled and will escape all of the usual schemes 
for detecting them.

In view of the more natural explanation of the ratio between R symmetry
and SUSY breaking scales, and the existence of a dark matter candidate 
in the bulk modulus scenario for SUSY breaking, we tentatively reject the
idea that SUSY breaking is triggered by the F term of a boundary modulus.
Its only advantage over the bulk modulus scenario is that we do not have
to massage coefficients of order one in order to push the reheat temperature 
above an MeV.  

For completeness, we should also discuss the possibility that SUSY breaking
itself is caused by gauge interactions which are weakly coupled at the 
fundamental scale.  This is required if we assume, with Dine 
\cite{mike}\cite{evadine}\cite{islands},
that moduli are fixed at some enhanced symmetry point.  Scenarios of
this sort are attractive because they allow us to use the idea of gauge
mediation \cite{fischlerdinenelson} to solve the SUSY flavor problem.
Gauge interactions generate superpotentials of the form
$\mu_1^3 {w_g}_1 (C_1/m_1) + \mu_2^3 {w_g}_2 (C_2/m_2)    $, where the
$C's$ are composite superfields and the $m_i$ the nonperturbative low energy
scales generated by asymptotic freedom.  Again, in order to cancel the
cosmological constant, we must introduce an R breaking gauge theory 
with scale $(m_1)$, 
which preserves SUSY and a SUSY breaking gauge theory, with scale
related by $m_1^6 = m_P^2 m_2^4$.   

There is no cosmological moduli problem in this picture, since all
moduli are assumed to be 
frozen by the initial superpotential.  Moduli and
dark matter in gauge mediated SUSY breaking models have been discussed in
\cite{cite}.

\subsection{Density Fluctuations Redux}

There is a small discrepancy in what we have said up till now, which
reader may have been rushed into ignoring.  We bragged about achieving
the right magnitude for energy density fluctuations of the inflaton, but
then proceeded to claim that the energy we see today in the universe
comes from another source entirely, {\it viz.} the true bulk moduli.

It is easy to see however that the true 
moduli inherit the fluctuations of the inflaton.  In a given region
the moduli fields start to oscillate when the Hubble constant is
about equal to their mass.  In a region of inflaton overdensity, this
will happen later and the ratio of modular energy density in the
overdense and average regions will start to increase like $a^3$.  This
will continue until the moduli in the overdense region begin to oscillate, a
fter which the ratio will remain constant.  Since the decrease of oscillating
modular energy and oscillating inflaton energy follows the same scaling
law, the magnitude of modular fluctuations will be the same as those in
the original inflaton field.  

We have assumed here that the true 
moduli begin to oscillate before the inflatons decay into radiation.
Since the reheat temperature is $10^9 $ GeV and the oscillation energy scale
is $10^{11}$ GeV, this assumption is valid.  

Another question to worry about is the possibility of large isocurvature
fluctuations in the true bulk moduli fields.  However, during inflation,
when the inflamoduli are excited away from their minimum, these are not
light fields.  The nonzero values of the inflamoduli break R symmetry.
The true moduli space is a ``river valley'' running
between the hills of the inflationary potential, and during inflation
the system lies in the hills above the valley, where the potential 
 is not flat in the valley direction.  Indeed, this situation persists
long into the era when the inflamoduli have begun to oscillate, because
of the factor of $10^{20}$ between the inflationary and SUSY violating
energy densities.   

These issues deserve a more careful analysis, because it is possible that
the transfer of fluctuations could leave some observable relic in the cosmic 
microwave background or that an observable level of isocurvature
fluctuations could be generated.  It is unclear to me whether reliable
conclusions can be obtained without more information about the nature of
the potential.  Nonetheless, it appears that to a first approximation,
the true moduli inherit the adiabatic perturbations of the inflaton
field, so that the estimates we made above can be directly related to
measurements of microwave background fluctuations.

\subsection{Generalizing Horava-Witten}

The moduli space of 11 dimensional SUGRA compactifications which preserve 
${\cal N} =1$ SUSY in four Minkowski dimensions splits into three components.
These are Joyce sevenfolds, F theory limits of compactification on Calabi-Yau
fourfolds, and Heterotic limits of compactification on $K3\times CY_3$.  
These may be continously connected when short distance physics 
is properly taken 
into account. In addition,
there may be many branches of moduli space which join onto these 
through generalized
extremal transitions.  The moduli space is thus highly complex.

The cosmological arguments of this paper indicate that the 
phenomenologically relevant
compactifications may belong to a highly constrained submanifold 
of this complicated space.
Namely, they should preserve eight supercharges in the bulk.  
The breaking to ${\cal N} =1$ 
should occur only on branes.  SUGRA compactifications preserving eight SUSYs are much
more constrained.  The holonomy must be contained in $SU(3)$ 
which implies that the
manifold is the product of a Calabi-Yau threefold times a torus, 
modded out by a discrete
group $\Gamma$.  In order to obtain a smooth manifold with eight SUSYs
, $\Gamma $ should act freely and the holonomy around the new cycles 
created by $\Gamma$
identification should be in $SU(3)$.  Clearly, a way to obtain 
Horava-Witten like
scenarios is to allow fixed manifolds of $\Gamma$, on which 
an additional SUSY is broken.
The original scenario of Horava and Witten was a 
$CY_3 \times S^1$ compactification in
which $\Gamma$ is a $Z_2$ reflection on the $S^1$.  The fixed planes 
carry $E8$ gauge
groups, and one must also choose an appropriate gauge bundle.  
A further generalization
allows five branes wrapped on two cycles of $CY_3$ to live between the planes.

It seems likely that more complicated choices of $\Gamma$ might 
lead to a wider class
of scenarios.  The problem of classifying scenarios of this type 
seems quite manageable
\footnote{Preliminary results on the classification problem have 
been obtained by
L.Motl.}.  The moduli space of 
compactifications of M theory on $CY_3$ times a torus has a reasonably
complicated structure, replete with extremal transitions.  Nonetheless, 
it is considerably
simpler than the fourfold or Joyce manifold problem, and we know much 
more about its
structure.  Thus, if cosmology really points us in the direction of 
generalized Horava-Witten
compactifications, we have made real progress in the search for the 
true vacuum of
M theory.

\section{Conclusions}

Witten's explanation of the discrepancy between the Planck and 
unification scales
in the context of Horava-Witten compactifications,
poses a challenge for inflationary cosmology and particularly for 
the notion that 
moduli are inflatons.  In fact, the enhanced bulk SUSY of these 
compactifications gives
us a clean definition of modular inflatons.  The scenario then makes an 
order of
magnitude prediction of the amplitude of primordial density fluctuations 
in terms
of the unification scale.

Cosmological arguments first discussed in \cite{preva} then focus 
attention on the
true moduli space of M theory, a locus of enhanced discrete R symmetry.  
Such a space
almost certainly exists \cite{bdmod}.  It is the attractor of postinflationary
cosmological evolution.  The further evolution of the universe then depends on 
whether this space contains bulk moduli.  In the attractive scenario 
in which it does,
the initial Hot Big Bang generated by inflation, is soon dominated by 
the energy
density stored in coherent oscillations of true bulk moduli.  
By making optimistic
but plausible assumptions about coefficients of order one in order of magnitude
estimates, one obtains a reheat temperature above that required 
by nucleosynthesis.
The decay of true bulk moduli, rather than that of the inflaton, 
generates the Hot Big
Bang of classical cosmology.  The baryon asymmetry must also be 
generated in these
decays, and this is possible if the SUSY standard model 
contains renormalizable baryon
number violating interactions (compatible with laboratory 
tests of baryon and lepton
number conservation).  As a consequence of this, there is no 
LSP dark matter candidate.
Instead, boundary moduli with a suppressed potential energy act as a natural
source of dark matter.  Indeed, the ratio between the Planck 
and unification scales
appears again in this scenario, this time in explaining the temperature at 
which matter and radiation make equal contributions to the energy 
density of the
Universe.  This estimate comes out an order of magnitude too high, 
but given the
crudity of the calculation it seems quite plausible that this mechanism 
could be
compatible with observation.  The \lq\lq dark modulus \rq\rq which 
appears in this
scenario could be a QCD axion with decay constant of order the
unification scale.
Our unconventional origin for the Hot Big Bang completely removes the 
cosmological upper
bound on this decay constant.  Such a particle would be undetectable in 
presently
proposed axion searches.

If a cosmology like that outlined here turns out to be correct, one 
might be tempted
to revise Einstein's famous estimate of the moral qualities of a 
hypothetical Creator.
The current standard model of cosmology was constructed in the sixties.  
Since then there
has been much speculation about cosmology at times earlier than that at 
which the
primordial elements were synthesized.  Most of it has been based on an 
eminently reasonable
extrapolation of the
Hot Big Bang to energy densities orders of magnitude higher.  
If the present scenario
is correct, no such extrapolation is possible, and the conditions 
in the Universe in
the first fraction of the First Three Minutes were considerably 
different from those 
at any subsequent time.  There was a prior Big Bang after 
inflation, whose remnants may be 
forever hidden from us.  The dark matter which dominates our 
universe is so weakly coupled
to ordinary matter that its detection is far beyond the reach of 
currently planned
experiments.   The QCD and electroweak phase transitions 
never occurred.

The only dramatic prediction of this scenario for currently 
planned experiments is
the occurrence of renormalizable baryon number violation 
in the low energy SUSY
world.  The details of the baryogenesis scenario envisaged here 
should be worked out
more carefully, and combined with laboratory constraints, 
to nail down precisely
which kind of operators are allowed.  The scenario is thus easily 
falsifiable, but even
the discovery of renormalizable baryon number violating 
interactions among SUSY particles 
will not be a confirmation of our cosmology.  Similarly, 
any evidence for the existence
of more or less conventional WIMP dark matter will be a 
strong indication that the present
speculations are incorrect, but the failure to discover 
WIMPS will not prove that they
are correct.

Instead one will have to rely on the slow accumulation of 
evidence against alternatives:
ruling out vanishing up quark mass and spontaneous CP violation 
as solutions to the strong
CP problem, the failure of conventional axion and 
WIMP searches, the discovery of
renormalizable B violation.  These will be steps on 
the road to proving that this cosmology
is correct, but the end of that road is not in sight.


\acknowledgments

I am  grateful to Sean Carroll, Willy Fischler, 
Lubos Motl and Paul Steinhardt for valuable discussions. This work 
was supported in part by the DOE under grant
number DE-FG02-96ER40559. 




\begin{thebibliography}{19}        

%

\bibitem{binetry} P.\,Binetruy, M.K.\,Gaillard,
{``Candidates For The Inflaton Field in Superstring Models,''}
\prd{34}{1986}{3069-3083}.
\bibitem{preva} T.\,Banks, M.\,Berkooz, S.H.\,Shenker,
G.\,Moore, P.J.\,Steinhardt,
{``Modular Cosmology,''}, \prd{52}{1995}{3548-3562}, \hepth{9503114};
T.\,Banks, M.\,Berkooz, P.J.\,Steinhardt,
{``The Cosmological Moduli Problem, Supersymmetry Breaking, And Stability
In Postinflationary Cosmology,''} \prd{52}{1995}{705-716},
\hepth{9501053}.


%
\bibitem{mike} M.\,Dine, {``Seeking the Ground State of String
Theory''}, Invited talk at 13th Nishinomiya-Yukawa Memorial Symposium on
Dynamics of Fields and Strings, Nishinomiya, Japan, 12-13 November,
1998. \hepth{9903212}; M.\,Dine, Y.\,Nir, Y.\,Shadmi, {``Enhanced
Symmetries and the Ground State of String Theory''}, Phys. Lett. B444,
103, (1998), \hepth{9806124}.
\bibitem{bdcoping} T.\,Banks, M.\,Dine, {`` Coping With Strongly Coupled
String Theory''}, Phys. Rev. D50, 7454, (1994) ,\hepth{9406132}.
\bibitem{bdaxion} T.\,Banks, M.\,Dine, {``Cosmology of String
Theoretic Axions''}, Nucl. Phys. B505, 445, (1997), \hepth{9608197}.
\bibitem{fischlerdinenelson} M.\,Dine, W.\,Fischler, {``A
Phenomenological Model of Particle Physics Based on Supersymmetry''},
Phys. Lett. 10B, 227, (1982): M.\,Dine, A.\,Nelson, {``Dynamical
Supersymmetry Breaking at Low Energies ''}, Phys. Rev. D48, 1277,
(1993), \hepph{9303230}.
\bibitem{othersandbks}  G.D.\,Coughlan, W.\,Fischler, E.W.\,Kolb,
S.\,Raby, G.G.\,Ross, {``Cosmological Problems for the Polonyi
Potential''}, Phys. Lett. 131B, 59, (1983); T.\,Banks, D.\,Kaplan,
A.\,Nelson, {``Cosmological Implications of Dynamical Supersymmetry
Breaking ''}, Phys. Rev. D49, 779, (1994), \hepph{9308292}.
\bibitem{bdmod} T.\,Banks, M.\,Dine, {``Quantum Moduli Spaces of N=1
String Theories''}, Phys. Rev. D53, 5790, (1996), \hepth{9508071}.
\bibitem{witsup} E.\,Witten, {``Nonperturbative Superpotentials in
String Theory''}, Nucl. Phys. B474, 343, (1996), \hepth{9604030}.
\bibitem{brustein} R.\,Brustein, P.J.\,Steinhardt, {``Challenges for
Superstring Cosmology''}, Phys. Lett. B302, 196, (1993), \hepth{9212049}.
\bibitem{dineseib} M.\,Dine, N.\,Seiberg, {``Is the Superstring Weakly
Coupled?''}, Phys. Lett. 162B, 299, (1985).
\bibitem{therminf} D.\,Lyth, E.\,Stewart, {``Thermal Inflation and the
Moduli Problem''}, Phys. Rev. D53, 1784, (1996), \hepph{9510204}.
\bibitem{irnir} L.\,Ibanez, G.G.\,Ross, {``Discrete Gauge Symmetry
Anomalies''}, Phys. Lett. B260, 291, (1991); G.\,Eyal, Y.\,Nir{``Lepton
Parity in Supersymmetric Models''}, \hepph{9904473}.
\bibitem{superK} T.\,Kajita (for the Super-Kamiokande and Kamiokande
Collaboration), {``Atmospheric Neutrino Results from Super Kamiokande
and Kamiokande: Evidence for Neutrino (mu) Oscillations''}, Talk given
at the 18th International Conference on Neutrino Physics and
Astrophysics, Takayama, Japan, 4-9 June 1998, \hepex{9810001}.
\bibitem{hrr} R.\,Holman, P.\,Ramond, G.G.\,Ross, Phys. Lett. 137B, 343,
(1984).
\bibitem{steve} S.\,Weinberg, {``Cosmological Production of Baryons''},
Phys. Rev. Lett. 42, 850, (1979).
\bibitem{wilczeketal} K.\,Babu, J.C.\,Pati, F.\,Wilczek, {``Fermion
Masses, Neutrino Oscillations, and Proton Decay in the Light of
SuperKamiokande''}, \hepph{9812538}
\bibitem{racetrack} L.\,Dixon, V.\,Kaplunosky, M.\,Peskin, unpublished;
N.V.\,Krasnikov, Phys. Lett. B193, 37, (1987). 
\bibitem{moore} J.\,Horne, G.\,Moore, {``Chaotic Coupling Constants''},
Nucl. Phys. B432, 109, (1994), \hepth{9403058}.
\bibitem{islands} A.\,Dabholkar, J.A.\,Harvey,
{``String Islands,''} \hepth{9809122}.
\bibitem{evadine} M.\,Dine, E.\,Silverstein,
{``New M-theory Backgrounds with Frozen Moduli,''}\newline
\hepth{9712166}.
\bibitem{horwita} P.\,Ho\hacek rava, E.\,Witten,
{``Heterotic and Type I String Dynamics from Eleven Dimensions,''}
\npb{460}{1996}{506-524}, \hepth{9510209}.
\bibitem{horwitb} E.\,Witten, {``Strong Coupling Expansion Of Calabi-Yau
Compactification,''} \npb{471}{1996}{135-158}, \hepth{9602070}.
\bibitem{strongcoup} T.\,Banks, M.\,Dine, 
{``Couplings And Scales In Strongly Coupled Heterotic String Theory,}
\npb{479}{1996}{173-196}, \hepth{9605136}.
\bibitem{cite} M.\,Kawasaki, N.\,Sugiyama, Modern Physics Letters A12,
1275, (1997), \hepph{9607273}; S.\,Dimopoulos, G.F.\,Giudice,
A.\,Pomarol, Phys. Lett. B389, 37. (1996), \hepph{9607225};
T.\,Asaka, M.\,Kawasaki, M.\,Yamaguchi, Phys. Lett. B451, 317, (1999), 
\hepph{9810334} (and references cited therein). 
%
\end{thebibliography}
\end{document}